\documentclass[aps,pra,floatfix,showpacs]{revtex4}

\usepackage[dvips]{graphicx}
\usepackage[english]{babel}
\usepackage{amsmath}
\usepackage{amssymb}
\usepackage{slashed}
\allowdisplaybreaks
\begin{document}
\title{Fundamental Constraints on Linear Response Theories of Fermi
Superfluids Above and Below $T_c$}

\author{Hao Guo$^{1,2}$,Chih-Chun Chien$^3$, Yan He$^{4,5}$, K. Levin$^5$}
\affiliation{$^1$Department of Physics, Southeast University, Nanjing 211189, China}
\affiliation{$^2$Department of Physics, University of Hong Kong, Hong Kong 999077, China}
\affiliation{$^3$Theoretical Division, Los Alamos National Laboratory, Los Alamos, NM, 87545, USA}
\affiliation{$^4$Department of Physics, University of California, Riverside}
\affiliation{$^5$James Franck Institute and Department of Physics,
University of Chicago, Chicago, Illinois 60637, USA}
\date{\today}


\begin{abstract}
We present fundamental constraints required for a consistent
linear response theory of fermionic superfluids and address temperatures both above
and below the transition temperature $T_c$. We emphasize two independent constraints, one
associated with gauge invariance (and the related Ward identity) and another
associated with the compressibility sum rule, both of which
are satisfied in strict BCS theory. However, we point out that it is the
rare many body theory which satisfies both of these. Indeed, well studied
quantum Hall systems and random-phase approximations to the electron gas are found
to have difficulties with meeting these constraints. We summarize two
distinct theoretical approaches which are, however, demonstrably compatible with
gauge invariance and the compressibility sum rule. The first of these
involves an extension of BCS theory to a mean field description of the BCS-Bose Einstein condensation
crossover. The second is the simplest Nozieres Schmitt-
Rink (NSR) treatment of pairing correlations in the normal state.
As a point of comparison we focus on the compressibility $\kappa$ of each and contrast
the predictions above $T_c$.
We note here that despite the compliance with sum rules, this NSR
based scheme leads to an unphysical divergence in $\kappa$ at
the transition.  Because of
the delicacy of the various consistency requirements, the results of this
paper suggest
that avoiding this divergence may repair one problem while at the same
time introducing others.
\end{abstract}

\maketitle

\section{Introduction: The Challenges of Arriving at Consistent Linear
Response Theories}

With recent progress in studies of ultracold Fermi superfluids undergoing
BCS-Bose Einstein condensation (BEC) crossover has
come a focus on generalized spin and charge susceptibilities (see, e.g., Ref.~\cite{Uedabook} for an introduction). These
in turn relate to the compressibility and spin susceptibility in the
linear response regime.
While strict BCS theory is well known to lead to fully consistent
results for linear response (see Ref.~\cite{OurPed} for a review), arriving at a generalization to address the
entire crossover is a major challenge.
The difficulty is to construct such theories as to be fully compatible
with $f$-sum and compressibility-sum rules which reflect conservation
principles. Although there have been some successes there are nevertheless
important failures, as we will address here.

In this paper, we discuss these challenges in the context of fermionic
superfluids.
We present examples of
theoretical approaches which are fully consistent with $f$-sum and
compressibility-sum rules. To our knowledge there
are two such theories beyond strict BCS theory and we discuss
both here.  The first of these involves a
mean-field approach to BCS-BEC crossover. Here we avoid the
complexities associated with pair fluctuation or
pseudogap effects \cite{OurAnnPhys}. The second of these involves
an investigation of the normal phase which includes pairing
fluctuations at the simplest level. In this context we address
Nozieres and Schmitt-Rink (NSR) \cite{NSR}
theory.
We demonstrate that gauge invariance
and sum rules can be made compatible with a linear response scheme.
Importantly, however,
this linear response theory is unphysical in that
it predicts a divergent
compressibility when approaching the superfluid transition temperature from
above.
This observation should underline the point noted above that a
treatment of linear response functions in fermionic superfluids
(and many-body systems overall) is full of pitfalls.

Indeed, for a general many-body theory, as
summarized in Ref.~\cite{Mahanbook}, it is difficult to obtain the same
result for the compressibility
via the derivatives of thermodynamic quantities as compared with
that via the two-particle correlation functions.
This issue has been extensively discussed for random-phase approximation (RPA) and RPA-generalizations
of the electron gas \cite{Singwi} as well as in quantum Hall systems \cite{HalperinQHE}.
From this literature,
it appears that when the two approaches for the compressibility disagree,
the more trustworthy scheme \cite{Singwi} is the thermodynamical approach.
In many ways more complex are the response functions of BCS theory
which from the very beginning \cite{Nambu60,Kadanoff61}
revealed difficulties associated with
incorporating charge conservation and gauge invariance in the broken-symmetry phase.

In a related fashion we note that there
are considerable discussions about where and when collective
mode contributions (which in neutral systems are phonon-like) associated with the fluctuations of the order parameter
 enter into
the electromagnetic response and related transport of fermionic
superfluids. These phononic modes are central to the bosonic superfluids.
Because correlation functions (and related transport coefficients)
are constrained
by sum rules, one is not free to incorporate these
collective modes in an arbitrary fashion.
In this paper we discuss in some detail how these collective modes
appear in a theory respecting conservation laws.
We will show that in the process of maintaining gauge invariance
one must self-consistently calculate the collective mode spectrum.
The usual phonons, which appear as
density waves above the transition, are
below $T_c$ strongly entangled with the phase and
(in general) amplitude modes of the order parameter.

Indeed, it appears that in the BCS superconductivity literature
(including helium-3 as the neutral counterpart),
phonons do not contribute to transport properties associated with
the transverse correlation functions.
This includes the conductivity, and shear viscosity.
A consequence of
this body of work based on
kinetic theory
as well as Kubo-based systematic studies
approaches \cite{Nozieres69,Stephentransport}
is that the normal fluid or condensate excitations involve
only the
fermionic quasi-particles.
In superfluid helium-3, as well, the normal fluid contributions to
thermodynamics reflect
only the fermionic quasi-particles.

This brings us to the possibility that the situation is different
for strong coupling superfluids in the sense of BCS-BEC crossover. Work by
our group showed
that within a generalized BCS-like description
the sum rules are satisfied without
including collective mode excitations contributing
to the normal fluid density \cite{OurPed}.
But there may be alternative theories where
at strong coupling the sound waves are important in this transport.
Indeed it has been argued \cite{Yu09}
 that these sound mode contributions are
important in the thermodynamics at unitarity, but it should
be noted that they were similarly invoked in
the BCS regime, in a manner which does not appear consistent with
theory and experiment on superfluid helium-3 \cite{WolfeHe3}.

Closely related to the charge and spin susceptibility are the dynamical
charge ($C$) and spin ($S$) structure factors: $S_{S},S_{C}$.
They are formally connected to the charge and spin response functions by the fluctuation-dissipation theorem \cite{Kubo_book}.
In a proper theory, the
conservation laws both for charge and spin yield well known
sum rule constraints on
$S_{S},S_{C}$.
In recent literature there has been an emphasis on these structure
factors
\cite{GiorginiCombescot,ValePRL,LittlewoodPRL05,SonThompson},
in part because they can be directly measured
via two photon Bragg experiments and in part, because they are thought
to reflect on an important parameter for the unitary gases: the so-called
``Contact parameter".

In this context it
has proved convenient to define
\begin{equation}
S_{\pm}(\omega,\mathbf{q}) \equiv \frac{1}{2}[S_{C} (\omega,\mathbf{q}) \pm
S_{S} (\omega,\mathbf{q})].
\end{equation}
%
%
It should be noted that in the literature \cite{GiorginiCombescot}
there is a tendency
to decompose these spin and charge structure factors
into separate spin components so that the density
(or spin) response is related to correlation functions
of the form
\begin{equation}
<[\rho_{\uparrow} \pm \rho_{\downarrow}]
[\rho_{\uparrow} \pm \rho_{\downarrow}]>
\label{eq:decomp}
\end{equation}
Presuming $S_{\uparrow \uparrow} = S_{\downarrow \downarrow}$ and
$S_{\uparrow \downarrow} = S_{\downarrow \uparrow}$ then one
infers
$$S(k, \omega) = 2 [S_{\uparrow \uparrow} ( k , \omega ) +
S_{\downarrow \uparrow} (k, \omega)]$$ with
$S_{\sigma \sigma'} = \sum_{n} <0|\rho_{\sigma}(k)|n>
<n|\rho_{\sigma'}^{\dagger} (k)|0> \delta(\omega - E_n)$.
In this way the difference structure factor
$S_{-}(\omega, \mathbf{q})$
is frequently associated with density correlations of
the form $<\rho_{\uparrow}~\rho_{\downarrow}>$.

This association derives from assuming that all
diagrams for the spin and charge response are equivalent and given
by the same combinations of charge density commutators (with only
simple sign changes) involving $\rho_{\uparrow}(r) \pm
\rho_{\downarrow}(r)$.
We emphasize in this paper that
this is specifically not the case below $T_c$ as a result of collective mode
effects which only couple to the density response function but decouple from the spin response function.
Ref.~\cite{OurPed} clearly demonstrates this difference for strict BCS theory.
Moreover, it needs not generally hold when
there is a different class of diagrams
required above $T_c$ in the spin and charge channels to insure
the $f$-sum rules.
While these crucial collective mode effects are sometimes inadvertently omitted
\cite{LittlewoodPRL05} in analyzing density-density correlation functions,
they are essential
for satisfying the longitudinal $f$-sum rule.

Of interest is a claim in
Ref.~\cite{ValePRL} that
the static structure factor (which involves an
integral over all frequencies) at large wavevector measures
the so-called Contact parameter. A rather different
observation was made by Son and Thompson \cite{SonThompson} who showed that
\textit{it is the high frequency, large wavevector structure factor
which is associated with the Contact}.
More precisely,
Son and Thompson
\cite{SonThompson}
investigated the relation between the structure factor and the
Contact interaction noting how delicate this issue is and that
``care should be taken not
 to
violate conservation laws".
This is the philosophy at the core of the present paper.

\section{
Superfluid Linear Response formalism} \label{C4}

We begin our discussion of linear response theory with the fundamental
Hamiltonian for a two-component Fermi gas interacting via contact interactions
\begin{equation}\label{H00}
H=\int d^3\mathbf{x}\psi^{\dagger}_{\sigma}(\mathbf{x})\Big(\frac{\hat{\mathbf{p}}^2}{2m}-\mu\Big)\psi_{\sigma}(\mathbf{x})-g\int
d^3\mathbf{\mathbf{x}}\psi^{\dagger}_{\uparrow}(\mathbf{x})\psi^{\dagger}_{\downarrow}(\mathbf{x})\psi_{\downarrow}(\mathbf{x})\psi_{\uparrow}(\mathbf{x}).
\end{equation}
We assume the interaction is attractive and $g$ is the bare coupling constant.
Here we adopt the convention $e=c=\hbar=1$ and the metric tensor
$g^{\mu\nu}$ is a diagonal matrix with the elements $(1,-1,-1,-1)$.

The goal of linear response theory is to find
the full electromagnetic (EM) vertex $\Gamma^{\mu}$ associated with the EM
response kernel
$ K ^{\mu\nu}(Q)$.
In the presence of a weak externally applied EM field with four-vector
potential $A^{\mu} = (\phi, \mathbf{A})$, the perturbed four-current density
$\delta J^{\mu}$ is given by
\begin{eqnarray}\delta J^{\mu}(Q) = K ^{\mu\nu}(Q) A_{\nu}(Q).
\end{eqnarray}
By introducing
the bare EM vertex
$\gamma^{\mu}(P+Q,P)=(1,\frac{\mathbf{p}+\frac{\mathbf{q}}{2}}{m})$ and full EM vertex $\Gamma^{\mu}(P+Q,P)$,
the gauge invariant EM response kernel can be expressed as
\begin{eqnarray}\label{KmnO}
K^{\mu\nu}(Q)=2\sum_{P}\Gamma^{\mu}(P+Q,P)G(P+Q)
\gamma^{\nu}(P,P+Q)G(P)+\frac{n}{m}h^{\mu\nu}, \end{eqnarray}
where $h^{\mu\nu}=-g^{\mu\nu}(1-g^{\nu0})$.
Throughout we define
$Q\equiv q^{\mu}=(i\Omega_{l},\mathbf{q})$ which is the 4-momentum of the external field with $\Omega_{l}$ being the boson Matsubara frequency, and $P\equiv p^{\mu}=(i\omega_n,\mathbf{p})$ is the 4-momentum of the fermion with $\omega_n$ being the fermion Matsubara frequency. $G(P)$ is the single-particle Green's function. The ``bare'' Green's function is given by $G_0(P)=(i\omega_n-\xi_{\mathbf{p}})^{-1}$ with $\xi_{\mathbf{p}}=\frac{p^2}{2m}-\mu$.

\subsection{Central constraints}

Gauge invariance and conservation laws impose an important
set of constraints on any linear response theory.
The full
EM vertex
$\Gamma^{\mu}$ must obey the Ward Identity \cite{OurPed}
\begin{eqnarray}\label{IWI}
q_{\mu}\Gamma^{\mu}(P+Q,P)=G^{-1}(P+Q)-G^{-1}(P)
\end{eqnarray}
and the identity associated with the compressibility
sum rule.
The former leads to the gauge invariant condition of the response kernel $q_{\mu}K^{\mu\nu}(Q)=0$ which further leads to the conservation of the perturbed current $q_{\mu}\delta J^{\mu}=0$. The compressibility sum rule imposes
an identity which we call the ``$Q$-limit Ward Identity " \cite{Maebashi09}
\begin{eqnarray}\label{QWI}
\lim_{\mathbf{q}\rightarrow\mathbf{0}}\Gamma^0(P+Q,P)|_{\omega=0}=\frac{\partial
 G^{-1}(P)}{\partial \mu}=1-\frac{\partial \Sigma(P)}{\partial \mu},
\end{eqnarray}
where $\Sigma(P)$ is the self energy, and the relation $G^{-1}(P)=G^{-1}_0(P)-\Sigma(P)$ has been applied. This will guarantee the compressibility sum rule $\partial n / \partial \mu=-K^{00}(0,\mathbf{q}\rightarrow\mathbf{0})$ with the compressibility given by $\kappa=n^{-2}(\partial n / \partial \mu)$, as will be shown shortly.

\subsection{Linear Response of BCS superfluids}

In BCS theory of fermionic superfluids, the order parameter is given by \begin{equation}\label{Delta}
\Delta(\mathbf{x})=g\langle\psi_{\downarrow}(\mathbf{x})\psi_{\uparrow}(\mathbf{x})\rangle.
\end{equation}
In the mean-field approximation, the Hamiltonian in the absence of
external fields may be written as
\begin{equation}\label{HBCS}
H=\int
d^3\mathbf{\mathbf{x}}\psi^{\dagger}_{\sigma}(\mathbf{x})\Big(\frac{\hat{\mathbf{p}}^2}{2m}
-\mu
\Big)\psi_{\sigma}(\mathbf{x})-\int
d^3\mathbf{\mathbf{x}}\Big(\Delta(\mathbf{x})\psi^{\dagger}_{\uparrow}(\mathbf{x})
\psi^{\dagger
}_{\downarrow}(\mathbf{x})+\textrm{h.c.}\Big).
\end{equation}
There are many reviews \cite{Schrieffer_book,Walecka} on how to
derive
$K ^{\mu\nu}(Q)$
at the BCS level. Here we set up an approach which we refer to
as the consistent fluctuation of the order parameter (or CFOP) theory.
Importantly, here, in contrast to the approach of Nambu \cite{Nambu60}
the changes in the phase and amplitude of the order parameter associated with
the external fields enter as additional components of the perturbation theory.
It is useful, however, to cast this CFOP theory in the Nambu
formulation. We remark that Ref.~\cite{Benfatto04} implemented an effective field theory 
for BCS superconductors and was able to obtained a gauge-invariant linear response theory.

We define
$\sigma_{\pm}=\frac{1}{2}(\sigma_1\pm i\sigma_2)$ where $\sigma_i$ is the Pauli matrix, and introduce
the Nambu-Gorkov spinors \begin{equation}\label{Ns}
\Psi_{\mathbf{p}}=\left[\begin{array}{c} \psi_{\uparrow\mathbf{p}} \\
\psi^{\dagger}_{\downarrow-\mathbf{p}}\end{array}\right], \qquad
\Psi^{\dagger}_{\mathbf{p}}=[\psi^{\dagger}_{\uparrow\mathbf{p}},\psi_{\downarrow-\mathbf{p}}].
\end{equation}
In the mean-field BCS approximation, the Hamiltonian (\ref{HBCS})
in the presence of electromagnetic (EM) fields can
be rewritten in the Nambu space as
\begin{eqnarray}\label{H0}
H=\sum_{\mathbf{p}}\Psi^{\dagger}_{\mathbf{p}}\xi_{\mathbf{p}}\sigma_3\Psi_{\mathbf{p}}+\sum_{\mathbf{p}\mathbf{q}}
\Psi^{\dagger}_{\mathbf{p}+\mathbf{q}}\big(-\frac{\mathbf{p}+\frac{\mathbf{q}}{2}}{m}\mathbf{A}_{\mathbf{q}}+\Phi_{\mathbf{q}}\sigma_3-\Delta_{\mathbf{q}}\sigma_+-\Delta^*_{-\mathbf{q}}\sigma_-\big)\Psi_{\mathbf{p}},
\end{eqnarray}
Here the order parameter is generalized to include fluctuations from its equilibrium value with
$\Delta_{\mathbf{q}}=g\sum_{\mathbf{p}}\langle\Psi^{\dagger}_{\mathbf{p}}\sigma_-\Psi_{\mathbf{p}+\mathbf{q}}\rangle$,
which will be imposed as a self-consistency condition. When the external EM field is
applied, the order parameter is perturbed and deviates from its equilibrium value.
The order parameter in equilibrium is $\Delta$, which is at $\mathbf{q}=\mathbf{0}$ and can be chosen to be real.
Denote the
small perturbation of the order parameter as $\Delta'_{\mathbf{q}}$ so
$\Delta_{\mathbf{q}}=\Delta+\Delta'_{\mathbf{q}}$. By introducing
$\Delta_{1\mathbf{q}}=-(\Delta'_{\mathbf{q}}+\Delta^{\prime\ast}_{-\mathbf{q}})/2$
and
$\Delta_{2\mathbf{q}}=-i(\Delta'_{\mathbf{q}}-\Delta^{\prime\ast}_{-\mathbf{q}})/2$,
the Hamiltonian (\ref{H0}) splits into two parts as $H=H_0+H'$ with one containing
the equilibrium quantities and the other containing the deviation from
equilibrium. \begin{eqnarray}\label{H1}
H_0=\sum_{\mathbf{p}}\Psi^{\dagger}_{\mathbf{p}}\hat{E}_{\mathbf{p}}\Psi_{\mathbf{p}},
\quad
H'=\sum_{\mathbf{p}\mathbf{q}}\Psi^{\dagger}_{\mathbf{p}+\mathbf{q}}\big(\Delta_{1\mathbf{q}}\sigma_1+\Delta_{2\mathbf{q}}\sigma_2+A_{\mu\mathbf{q}}\hat{\gamma}^{\mu}(\mathbf{p}+\mathbf{q},\mathbf{p})\big)\Psi_{\mathbf{p}},
\end{eqnarray}
where $\hat{E}_{\mathbf{p}}=\xi_{\mathbf{p}}\sigma_3-\Delta\sigma_1$
is an energy operator and
$\hat{\gamma}^{\mu}(\mathbf{p}+\mathbf{q},\mathbf{p})=(\sigma_3,\frac{\mathbf{p}+\frac{\mathbf{q}}{2}}{m})$
 is the bare EM vertex in the Nambu space.
Here the perturbation of the order parameter and the EM perturbation are treated on
 equal footing and this will
naturally lead to gauge invariance of the linear response theory.
 The quasi-particle energy is given by
$E_{\mathbf{p}}=\sqrt{\xi^2_{\mathbf{p}}+\Delta^2}$. The propagator in the Nambu space is
\begin{eqnarray}\label{NGP}
\hat{G}(P)\equiv\hat{G}_{\mathbf{p}}(i\omega_n)=\frac{1}{i\omega_n-\hat{E}_{\mathbf{p}}}=\left(\begin{array}{cc}G(P)
& F(P)\\F(P) & -G(-P)\end{array}\right), \end{eqnarray}
where
\begin{eqnarray}\label{Green}
G(P)=\frac{u^2_{\mathbf{p}}}{i\omega_n-E_{\mathbf{p}}}+\frac{v^2_{\mathbf{p}}}{i\omega_n+E_{\mathbf{p}}},\quad
F(P)=-u_{\mathbf{p}}v_{\mathbf{p}}\Big(\frac{1}{i\omega_n-E_{\mathbf{p}}}-\frac{1}{i\omega_n+E_{\mathbf{p}}}\Big)
\end{eqnarray} are the single-particle Green's function and anomalous Green's
function respectively and
$u^2_{\mathbf{p}},v^2_{\mathbf{p}}=\frac{1}{2}(1\pm\frac{\xi_{\mathbf{p}}}{E_{\mathbf{p}}})$.

By introducing the generalized driving potential and generalized interacting vertex
 \begin{eqnarray}\label{GGPS}
\hat{\mathbf{\Phi}}_{\mathbf{q}}=\big(\Delta_{1\mathbf{q}},\Delta_{2\mathbf{q}},A_{\mu\mathbf{q}}\big)^T,\quad
\hat{\mathbf{\Sigma}}(\mathbf{p}+\mathbf{q},\mathbf{p})=\big(\sigma_1,\sigma_2,\hat{\gamma}^{\mu}(\mathbf{p}+\mathbf{q},\mathbf{p})\big)^T,
\end{eqnarray}
the generalized perturbed current
$\vec{\eta}$ is given by \begin{eqnarray}\label{RF1}
\vec{\eta}(\tau,\mathbf{q})=\sum_{\mathbf{p}}\langle\Psi^{\dagger}_{\mathbf{p}}(
\tau)\hat{\mathbf{\Sigma}}(\mathbf{p}+\mathbf{q},\mathbf{p})\Psi_{\mathbf{p}+
\mathbf{q}}(\tau)\rangle+\frac{n}{m}\delta^{i3}h^{\mu\nu}A_{\nu}(\tau,\mathbf{q}).
\end{eqnarray} where the component 
$\eta^{\mu}_3=\langle J^{\mu}\rangle$ denotes the EM current
and $\eta_{1,2}$ the perturbations of the gap function.
This leads to a linear response equation
in a matrix form \begin{eqnarray}\label{eqn:Q}
\vec{\eta}(\omega,\mathbf{q})&=&\tensor{Q}(\omega,\mathbf{q})\cdot\hat{\mathbf{\Phi}}
(\omega,\mathbf{q})
\nonumber \\ &=&\left( \begin{array}{ccc} Q_{11}(\omega,\mathbf{q}) &
Q_{12}(\omega,\mathbf{q}) & Q^{\nu}_{13}(\omega,\mathbf{q}) \\
Q_{21}(\omega,\mathbf{q}) & Q_{22}(\omega,\mathbf{q}) &
Q^{\nu}_{23}(\omega,\mathbf{q}) \\ Q^{\mu}_{31}(\omega,\mathbf{q}) &
Q^{\mu}_{32}(\omega,\mathbf{q}) &
Q^{\mu\nu}_{33}(\omega,\mathbf{q})+\frac{n}{m}h^{\mu\nu} \end{array}\right)\left(
\begin{array}{ccc} \Delta_{1}(\omega,\mathbf{q}) \\ \Delta_{2}(\omega,\mathbf{q})
 \\A_{\nu}(\omega,\mathbf{q})
\end{array}\right). \end{eqnarray}
The response functions $Q_{ij}$ are
\begin{eqnarray} Q_{ij}(\tau-\tau',\mathbf{q})=-\sum_{\mathbf{p}\mathbf{p}'}\langle
T_{\tau}[\Psi^{\dagger}_{\mathbf{p}}(\tau)\hat{\Sigma}_i(\mathbf{p}+\mathbf{q},\mathbf{p})\Psi_{\mathbf{p}+\mathbf{q}}(\tau)\Psi^{\dagger}_{\mathbf{p}'+\mathbf{q}}(\tau')\hat{\Sigma}_j(\mathbf{p}',\mathbf{p}'+\mathbf{q})\Psi_{\mathbf{p}'}(\tau')]\rangle.
\end{eqnarray}
Using the Wick decomposition \cite{Walecka}, we obtain
\begin{eqnarray}\label{eqn:RF} Q_{ij}(i\Omega_{l},
\mathbf{q})=\textrm{Tr}T\sum_{i\omega_n}\sum_{\mathbf{p}}
\big(\hat{\Sigma}_i(P+Q,P)\hat{G}(P+Q)\hat{\Sigma}_j(P,P+Q)\hat{G}(P)\big),
\end{eqnarray}
The gap
equation leads to $\eta_{1,2}=-\frac{2}{g}\Delta_{1,2}$. 
and using Eq.(\ref{eqn:Q}), we find \begin{eqnarray}\label{D1D2}
\Delta_1=-\frac{Q^{\nu}_{13}\tilde{Q}_{22}-Q^{\nu}_{23}Q_{12}}{\tilde{Q}_{11}\tilde{Q}_{22}-Q_{12}Q_{21}}A_{\nu},\quad\Delta_2=-\frac{Q^{\nu}_{23}\tilde{Q}_{11}-Q^{\nu}_{13}Q_{21}}{\tilde{Q}_{11}\tilde{Q}_{22}-Q_{12}Q_{21}}A_{\nu}.
\end{eqnarray} where $\tilde{Q}_{11}\equiv \frac{2}{g}+Q_{11}$ and
$\tilde{Q}_{22}\equiv \frac{2}{g}+Q_{22}$.

The quantity of interest is the EM response kernel $K^{\mu\nu}$, which has the following form in
the Nambu space.
\begin{eqnarray}\label{Kmn1}
K^{\mu\nu}(Q)=\textrm{Tr}\sum_P\big(\hat{\Gamma}^{\mu}(P+Q,P)\hat{G}(P+Q)\hat{\gamma}^{\nu}(P,P+Q)\hat{G}(P)\big)+\frac{n}{m}h^{\mu\nu}.
\end{eqnarray}
After substituting Eq.~(\ref{D1D2}) into our linear response
expression we find
\begin{eqnarray}\label{CEtemp}
J^{\mu}=Q^{\mu}_{31}\Delta_1+Q^{\mu}_{32}\Delta_2+(Q^{\mu\nu}_{33}+\frac{n}{m}h^{\mu\nu})A_{\nu},
\end{eqnarray} from which we obtain
\begin{eqnarray}\label{dK} K^{\mu\nu}=\tilde{Q}^{\mu\nu}_{33}+\delta
K^{\mu\nu},\quad \delta
K^{\mu\nu}=-\frac{\tilde{Q}_{11}Q^{\mu}_{32}Q^{\nu}_{23}+\tilde{Q}_{22}Q^{\mu}_{31}Q^{\nu}_{13}-Q_{12}Q^{\mu}_{31}Q^{\nu}_{23}-Q_{21}Q^{\mu}_{32}Q^{\nu}_{13}}{\tilde{Q}_{11}\tilde{Q}_{22}-Q_{12}Q_{21}}.
\end{eqnarray} Here $\tilde{Q}^{\mu\nu}_{33}=Q^{\mu\nu}_{33}+\frac{n}{m}h^{\mu\nu}$. Hence the effects of fluctuations of the order parameter are included in the response kernel. From the expression of $K^{\mu\nu}$, the full EM vertex $\hat{\Gamma}^{\mu}(P+Q,P)$ in the Nambu space can be determined. We define
\begin{eqnarray}\label{tmp4}
\Pi^{\mu}_1=\frac{\left|\begin{array}{cc}Q^{\mu}_{31} & Q_{21}\\ Q^{\mu}_{32} &
\tilde{Q}_{22}\end{array}\right|}{\left|\begin{array}{cc}\tilde{Q}_{11} & Q_{12}\\
Q_{21} & \tilde{Q}_{22}\end{array}\right|}, \mbox{
}\Pi^{\mu}_2=\frac{\left|\begin{array}{cc}Q^{\mu}_{32} & Q_{12} \\ Q^{\mu}_{31} &
\tilde{Q}_{11} \end{array}\right|}{\left|\begin{array}{cc}\tilde{Q}_{11} & Q_{12}\\
Q_{21} & \tilde{Q}_{22}\end{array}\right|}. \end{eqnarray}
From Eq.~(\ref{dK}), $K^{\mu\nu}$ can be expressed as
\begin{eqnarray}\label{Kmn2}
K^{\mu\nu}(Q)
&=&\textrm{Tr}\sum_P\big([\hat{\gamma}^{\mu}(P+Q,P)-\sigma_1\Pi_1^{\mu}(Q)-\sigma_2\Pi_2^{\mu}(Q)]\hat{G}(P+Q)\hat{\gamma}^{\nu}(P,P+Q)\hat{G}(P)\big)+\frac{n}{m}h^{\mu\nu}.
\end{eqnarray}
where we have used
Eq.~(\ref{eqn:RF}) to arrive at $Q^{\mu\nu}_{33}$, $Q^{\nu}_{13}$ and $Q^{\nu}_{23}$. One can further identify 
\begin{eqnarray}\label{FEM}
\hat{\Gamma}^{\mu}(P+Q,P)=\hat{\gamma}^{\mu}(P+Q,P)-\sigma_1\Pi^{\mu}_1(Q)-\sigma_2\Pi^{\mu}_2(Q).
\end{eqnarray}

Our ultimate goal will be to demonstrate consistency with
the various Ward identities within this BCS formulation.
In order to proceed we rewrite
Eq.~(\ref{dK})
in the form of
Eq.~(\ref{KmnO}).
For this purpose it
is convenient to define
\begin{equation}\label{OWI1}
\Pi^{\mu}(Q)=-\Pi^{\mu}_1(Q)+i\Pi^{\mu}_2(Q),\quad
\bar{\Pi}^{\mu}(Q)=-\Pi^{\mu}_1(Q)-i\Pi^{\mu}_2(Q), \end{equation}

With straightforward algebraic manipulations one arrives at
\begin{eqnarray}\label{Pmunu1}
K^{\mu\nu}(Q)&=&2\sum_P\big[\gamma^{\mu}(P+Q,P)G(P+Q)\gamma^{\nu}(P,P+Q)G(P)+\Pi^{\mu}(Q)F(P+Q)\gamma^{\nu}(P,P+Q)G(P)\nonumber\\
&+&\bar{\Pi}^{\mu}(Q)G(P+Q)\gamma^{\nu}(P,P+Q)F(P)-\gamma^{\mu}(-P,-P-Q)F(P+Q)\gamma^{\nu}(P,P+Q)F(P)\big]+\frac{n}{m}h^{\mu\nu}.
\end{eqnarray}
Similar expressions for the density response function has also been obtained using a kinetic-theory approach \cite{Stringari06}.
Substituting the expression $F(P)=\Delta G_0(-P)G(P)$ into
Eq.(\ref{Pmunu1}), and then comparing with Eq.(\ref{KmnO}), we arrive at an
expression for the full EM
vertex \begin{eqnarray}\label{Gamma1}
\Gamma^{\mu}(P+Q,P)&=&\gamma^{\mu}(P+Q,P)+\Delta\Pi^{\mu}(Q)G_0(-P-Q)+\Delta\bar{\Pi}^{\mu}(Q)G_0(-P)\nonumber\\
& &-\Delta^2G_0(-P)\gamma^{\mu}(-P,-P-Q)G_0(-P-Q), \end{eqnarray}
While rather complex, this represents an important result. The second and
third terms correspond to the contributions associated with collective-mode effects \cite{OurPed},
while the fourth term can be identified with the so-called Maki-Thompson diagram \cite{OurAnnPhys}.

We remark that in the standard calculation of the Meissner effect and superfluid density \cite{Walecka} the collective-mode contribution is not included. This is because the current-current correlation functions are evaluated and they correspond to the transverse components of Eq.~\eqref{Pmunu1}. One can show that the collective-mode contribution cancel in the limit $\omega=0,\mathbf{q}\rightarrow\mathbf{0}$ \cite{OurPed} so the collective-mode effect may be ignored. In contrast, we will show in the following that the collective-mode effect for longitudinal response
functions is crucial in restoring gauge invariance.

\section{Verification of Self Consistency in Linear Response}

In this section we demonstrate that the Ward identities and the
$Q$-limit Ward identity (associated with the compressibility sum rule)
are consistently satisfied at this mean field level.
We first present arguments to show that
the vertex
function in Eq.~\eqref{Gamma1}
obeys the Ward Identity (\ref{IWI}).
Contracting both sides of Eq.(\ref{Gamma1}) with $q^{\mu}$, we
have \begin{eqnarray}
q_{\mu}\Gamma^{\mu}(P+Q,P)&=&G_0^{-1}(P+Q)-G_0^{-1}(P)-2\Sigma(P+Q)+2\Sigma(P)-\frac{\Sigma(P+Q)\Sigma(P)}{\Delta^2}\big(G_0^{-1}(-P)-G_0^{-1}(-P-Q)\big)\nonumber\\
&=&G^{-1}(P+Q)-G^{-1}(P), \end{eqnarray} which is the desired
result. Here we have used the fact
that $\Sigma(P)=-\Delta^2G_0(-P)$ for BCS superfluids \cite{Ourreview}.
It can
be proved analytically that gauge invariance implies that
the density-density response function always satisfies
the $f$-sum rule (for details, see Ref.\cite{OurPed})
\begin{eqnarray}\label{eqn:fS}
\int^{\infty}_0d\omega\omega \chi_{\rho\rho}(\omega,\mathbf{q})=n\frac{q^2}{2m}.
\end{eqnarray}
Here we define the density-density correlation function
$$\chi_{\rho\rho}=-\frac{1}{\pi}\textrm{Im}K^{00}.$$

We turn now to the $Q$-limit Ward identity from which
the compressibility sum rule can be derived \cite{Maebashi09}.
We note that $G^{-1}(P)=G^{-1}_0(P)-\Sigma(P)$, so that
\begin{eqnarray}\label{CSR}
\frac{\partial n}{\partial \mu}&=&2\sum_P\frac{\partial G(P)}{\partial \mu}=-2\sum_PG^2(P)\frac{\partial G^{-1}(P)}{\partial \mu}\nonumber\\
&=&-2\sum_PG^2(P)\Big(1-\frac{\partial \Sigma(P)}{\partial \mu}\Big)=-2\sum_P\Gamma^0(P,P)G(P)\gamma^0(P,P)G(P)\nonumber\\
&=&-K^{00}(\omega=0,\mathbf{q}\rightarrow\mathbf{0}),
\end{eqnarray}
where in the last line, the expression in Eq.~(\ref{KmnO})
has been applied.
This analysis demonstrates that the compressibility obtained via thermodynamic
arguments relates to properties of two particle correlation functions.

The more explicit
proof of the $Q$-limit Ward identity (\ref{QWI}) is briefly outlined here.
Since $\lim_{\mathbf{q}\rightarrow\mathbf{0}}G_0(P+Q)|_{\omega=0}=G_0(P)$, we evaluate
$\Gamma^0(P+Q,P)$ in the limit $\omega=0$ and $\mathbf{q}\rightarrow\mathbf{0}$:
\begin{eqnarray}\label{LHS}
\lim_{\mathbf{q}\rightarrow\mathbf{0}}\Gamma^0(P+Q,P)|_{\omega=0}
&=&1-2\Delta\lim_{\mathbf{q}\rightarrow\mathbf{0}}\Pi^{0}_1(Q)|_{\omega=0}G_0(-P)-\Delta^2G^2_0(-P).
\end{eqnarray} Using $\Sigma(P)=-\Delta^2G_0(-P)$, the right hand side of Eq.~(\ref{QWI})  
is
\begin{eqnarray}\label{RHS} 1-\frac{\partial \Sigma(P)}{\partial
\mu}=1+2\Delta\frac{\partial \Delta}{\partial \mu}G_0(-P)-\Delta^2G^2_0(-P),
\end{eqnarray} where the identity
$\partial_{\mu}G_0(-P)=-G^2_0(-P)\partial_{\mu}G^{-1}_0(-P)=-G^2_0(-P)$ has been
applied. Comparing Eqs.(\ref{LHS}) and (\ref{RHS}), one can see that the
$Q$-limit Ward identity holds for BCS theory only when \begin{eqnarray}\label{t1}
\frac{\partial \Delta}{\partial
\mu}=-\lim_{\mathbf{q}\rightarrow\mathbf{0}}\Pi^{0}_1(Q)|_{\omega=0}. \end{eqnarray}
The left hand side can be evaluated by differentiating both sides of the gap equation $1=\frac{1}{\Delta}\sum_PF(P)$ with respect to $\mu$.
\begin{eqnarray}\label{pDpm}
\frac{\partial\Delta}{\partial\mu}=\frac{\sum_{\mathbf{p}}\frac{\xi_{\mathbf{p}}}{E^2_{\mathbf{p}}}\Big(\frac{1-2f(E_{\mathbf{p}})}{E_{\mathbf{p}}}+2\frac{\partial
f(E_{\mathbf{p}})}{\partial
E_{\mathbf{p}}}\Big)}{\sum_{\mathbf{p}}\frac{\Delta}{E^2_{\mathbf{p}}}\Big(\frac{1-2f(E_{\mathbf{p}})}{E_{\mathbf{p}}}+2\frac{\partial
f(E_{\mathbf{p}})}{\partial
E_{\mathbf{p}}}\Big)}.
\end{eqnarray}
By using the expressions of the response functions given in the Appendix, one can show that
\begin{eqnarray}\label{pDpm2}
\frac{\partial\Delta}{\partial\mu}=-\frac{Q^0_{13}(0,\mathbf{q}\rightarrow\mathbf{0})}{\tilde{Q}_{11}(0,\mathbf{q}\rightarrow\mathbf{0})}=-\lim_{\mathbf{q}\rightarrow\mathbf{0}}\Pi_1^0(Q)\Big|_{\omega=0}.
\end{eqnarray}
Thus the $Q$-limit Ward identity is respected, which then guarantees
the compressibility sum rule.

It is of interest to address
why an RPA-based approach usually fails 
to satisfy the compressibility sum rule \cite{Singwi,Mahanbook}. In the RPA approach, one starts with the expression of the density susceptibility called $(\partial n/\partial \mu)_0$ from thermodynamics or from a simpler model. By introducing a summation over a series of bubble diagrams, the RPA approach leads to the expression $(\partial n/\partial \mu)_{RPA}=(\partial n/\partial \mu)_0/[1-\lambda(\partial n/\partial \mu)_0]$, where $\lambda$ is a combination of the coupling constant representing
bubble diagrams and the combinatoric factors for counting the series \cite{Mahanbook}. For finite values of $(\partial n/\partial \mu)_0$ and $\lambda$, the RPA result always differs from $(\partial n/\partial \mu)_0$ and this makes it particularly difficult to satisfy the compressibility sum rule, which requires a consistent expression for $\partial n/\partial \mu$.

\section{Application to BCS-BEC Crossover at the mean-field level}
\label{CS7}

We make the important observation that
the arguments presented above for the self consistency of linear
response in strict BCS theory can be readily extended to treat
BCS-BEC crossover theory \textit{at the mean field level}.
This level is, of course,
not fully adequate because it does not include fluctuations associated
with non-condensed pairs. Nevertheless, it does provide an example
of a fully sum-rule consistent approach to linear response in
fermionic superfluids.

At this mean field level, the
number density and the order parameter in equilibrium can be written as
a function of temperature in terms of the equations

\begin{eqnarray}\label{NGE}
n=\sum_{\mathbf{p}}\Big[1-\frac{\xi_{\mathbf{p}}}{E_{\mathbf{p}}}\big(1-2f(E_
{\mathbf{p}})\big)\Big],
\quad \frac{1}{g}&=&\sum_{\mathbf{p}}\frac{1-2f(E_{\mathbf{p}})}{2E_{\mathbf{p}}
}.
\end{eqnarray}
Because we contemplate arbitrary chemical potentials away from the
BCS regime where $\mu \equiv E_F$, this theory applies
to the
whole BCS-BEC crossover regime.
Important here is the introduction of the two body scattering length
$a$ via a renormalization of the coupling constant \cite{Leggett,largeNcrossover}
\begin{eqnarray} \frac{1}{g}=\frac{m}{4\pi
a}-\sum_{\mathbf{p}}\frac{1}{2\epsilon_{\mathbf{p}}}, \end{eqnarray}
where $a$ is the $s$-wave
scattering length and $\epsilon_{\mathbf{p}}=p^2/2m$. In the BCS limit, $a\rightarrow
0^-$, $\Delta\rightarrow0$ and $\mu\rightarrow E_F$ where $E_F$ is the Fermi energy.
While in the deep BEC regime, $a\rightarrow 0^+$, $\Delta\rightarrow\infty$ and
$\frac{\Delta}{|\mu|}\rightarrow 0$.

\begin{figure}[tb]
\includegraphics[width=5.5in,clip]
{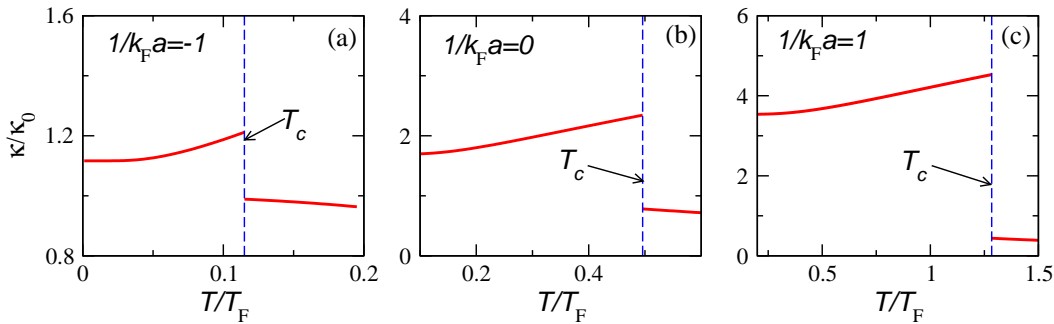}
\caption{Compressibility at the strict mean field level and
as function of temperature from BCS to BEC. $T_F$ is the Fermi temperature. $\kappa_0=\frac{3}{2}\frac{1}{nE_F}$ is the compressibility for a noninteracting Fermi gas at $T=0$.}
\label{fig:C}
\end{figure}

One of the best measures of a linear response theory is the calculation
of the compressibility
$$\kappa=n^{-2}(dn/d\mu)$$
based on the density correlation functions.
This is particularly problematic because of the difficulty of
finding the same answer as found from thermodynamics.
It is of considerable interest, then,
to establish the form of the compressibility
in a theory with full (compressibility) sum rule compatibility. At the
mean field level the behavior above the transition temperature is that of a free
Fermi gas.
Below $T_c$, the
compressibility either via thermodynamics
or via the two body density density response
leads to
\begin{eqnarray}\frac{\partial n}{\partial \mu}=-K^{00}(0,\mathbf{q}\rightarrow\mathbf{0})
\end{eqnarray}
 where
$K^{00}=Q^{00}_{33}+\delta K^{00}$.

Fig.\ref{fig:C} plots the compressibility $\kappa$ as a function of
temperature for $1/k_Fa=-1$ (on the BCS side), $0$ (the unitary point), and $1$ (on the BEC side). The figure exhibits an expected thermodynamic
signature of a phase transition, appearing as a discontinuity
in the compressibility at $T_c$. The discontinuity in $\kappa$
at $T_c$ can be traced back to
the appearance of collective-mode term $\delta K^{00}$ which
sets in below $T_c$ and is absent in the normal state.
It should be noted that
at $T_c$, $\delta K^{00}$ is finite only when $\omega=0$. When $\omega$ approaches but
does not equal $0$, we have $\delta K^{00}\rightarrow0$ at $T_c$. In this
way
$\delta K^{00}$ is not analytic at $\omega=0$.

Additional properties of the BCS to BEC crossover can be analyzed
similarly. For example, in
Appendix~\ref{app:b}
we
analytically evaluate the $T=0$ density
structure factor in the BCS and BEC limits.
Here one sees that at $T=0$ the density structure factor for low frequency and momentum is dominated by the gapless collective mode. As one crosses from BCS to BEC, this mode appears as the
usual sound mode of BCS theory and
evolves continuously into the Bogoliubov mode and eventually to the free bosonic
dispersion in the BEC regime.

One can similarly address the spin response functions following,
for example the derivation in Ref.~\cite{OurPed}.
Importantly,
the collective modes appear only in the density
response and
do not couple to the spin response functions.
This
supports the discussion given in the introduction
that any algebra involving both the density and spin response functions which
decomposes these functions into separate $\uparrow$
and $\downarrow$ contributions (see Eq.~(\ref{eq:decomp}))
is generally problematic, except in the absence of interactions.
Such algebraic manipulations are not possible when the
diagram sets in different channels are not the same.

\section{Consistent Linear Response Theory Above $T_c$: Example of Pair
Correlated State}
\label{CS20}

We now turn to the compressibility in a theory (of the normal phase)
which includes pair correlations. It is notable that here too, one
finds consistency with the usual Ward (and $Q$-limit Ward) identities,
providing one restricts consideration to the theory originally
introduced by Nozieres and Schmitt-Rink \cite{NSR} for the normal phase only.
The NSR paper was among the first to emphasize
the importance of treating
pair correlations in the normal phase. Indeed these were discussed
along with an analysis of
the ground state considered here and introduced by
Leggett \cite{Leggett} and Eagles \cite{Eagles}. Interestingly,
theories
which incorporate correlated pairs which are
based on this NSR scheme do \textit{not} appear to relate to the BCS-Leggett
ground state \cite{OurAnnPhys}.
This is, in part a reflection of the rather ubiquitous first order
transition associated with extending NSR theory below $T_c$.
Our group \cite{Ourreview}
has extensively discussed one approach which
appears (rather uniquely) to lead to a second order transition from
a different (as compared to NSR) normal phase into this well known ground state.
However, it is more complicated than the NSR theory, and the related
compressibility will be presented elsewhere.

Here, in order to illustrate a fully consistent approach to
linear response in the normal phase (beyond that of a noninteracting Fermi gas)
we use the simpler NSR scheme.
Even though it has been improved and reviewed many times (see Refs.~\cite{Pethickbook,Uedabook} for reviews), a full
discussion on the linear response theory within NSR theory is still lacking. In a previous publication \cite{OurAnnPhys} we have shown that by
carefully choosing a set of diagrams for the vertex function, the NSR theory
respects the usual Ward Identity associated with
gauge invariance.  Here we will show that the compressibility
derived from this
vertex function also satisfies the compressibility sum rule.

\begin{figure}
\centerline{\includegraphics[clip,width=3.4in]{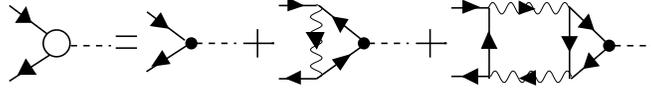}}
\caption{\label{fig:NSRvertex} The diagrams for the vertex function of NSR theory. The first one on the right hand side is the bare vertex, the second one is the ``MT'' diagram, and the last one is the ``AL'' diagram. Hollow and solid dots denote full and bare vertices. Solid lines and wavy lines correspond to propagator of non-interacting fermions and t-matrix, respectively. Due to the two ways of connecting the fermion propagator inside a $t$-matrix, there are two AL diagrams.}
\end{figure}

We begin with a brief review of the NSR theory and its linear response theory. The self energy is $\Sigma(K)=\sum_{Q} t_0(Q)G_0(Q-K)$ with $t_0(Q)=1/[g^{-1}+\chi_0(Q)]$ being the $t$-matrix in which $\chi_0(Q)=\sum_{P} G_0(P)G_0(Q-P)$ is the pair susceptibility. The number equation is given by $n=2\sum_{K}G(K)$ while an approximate number equation was implemented in the original NSR paper \cite{NSR}. The response function may also be written formally as
 \begin{eqnarray}\label{QmnG2}
K^{\mu\nu}(i\Omega_l,\mathbf{q})=2\sum_{P}\Gamma^{\mu}(P+Q,P)G(P+Q) \gamma^{\nu}(P,P+Q)G(P)+\frac{n}{m}h^{\mu\nu},
\end{eqnarray}
where the full EM vertex function $\Gamma^{\mu}$ must obey the Ward identity (\ref{IWI})
so that $q_{\mu}K^{\mu\nu}(Q)=0$.
The correction to the full vertex function should be consistent with that of the self energy, hence it
is associated with the set of diagrams shown in Fig.\ref{fig:NSRvertex}.
We have
\begin{eqnarray}\label{VNSR}\Gamma^{\mu}(P+Q,P)=\gamma^{\mu}(P+Q,P)+\textrm{MT}^{\mu}(P+Q,P)+\textrm{AL}^{\mu}(P+Q,P),\end{eqnarray}  where we
identify the Maki-Thompson (MT) and Aslamazov-Larkin (AL) diagrams with
\begin{eqnarray}\label{eq:G0G0eq}
\textrm{MT}^{\mu}(P+Q,P) &=& \sum_{K}t_0(K)G_{0}(K-P)\gamma^{\mu}(K-P,K-P-Q)G_{0}(K-P-Q), \nonumber \\
\textrm{AL}^{\mu}(P+Q,P) &=& -2\sum_{L,K}t_0(K)t_0(K+Q)G_{0}(K-P)G_{0}(K-L)G_{0}(L+Q)\gamma^{\mu}(L+Q,L)G_{0}(L). \nonumber
\end{eqnarray}
The factor $2$ in the AL diagram comes from the fact that the vertex can
be inserted in one of the two fermion propagators in the $t$-matrix and the minus sign is because inserting the vertex splits the t-matrix in the self energy.

To prove that
the full vertex in Eq.~(\ref{VNSR}) satisfies the Ward Identity we
contract the
MT and AL terms with $q_{\mu}$ to yield
\begin{eqnarray}
q_{\mu} \textrm{MT}^{\mu}(P+Q,P)&=&\sum_{K}t_0(K)G_{0}(K-P)[G_{0}^{-1}(K-P)-G_{0}^{-1}(K-P-Q)]G_{0}(K-P-Q), \nonumber \\
&=& -[\Sigma(P)-\Sigma(P+Q)]. \\
q_{\mu}\textrm{AL}^{\mu}(P+Q,P) &=&-2\sum_{L,K}t_0(K)t_0(K+Q)G_{0}(K-P)G_{0}(K-L)G_{0}(L+Q)[G_{0}^{-1}(L+Q)-G_{0}^{-1}(L)]G_{0}(L), \nonumber \\
&=& 2[\Sigma(P)-\Sigma(P+Q)].
\end{eqnarray}
In deriving the second relation, the identity
$\chi_0(K)-\chi_0(K+Q)=t_0^{-1}(K)-t_0^{-1}(K+Q)$ has been applied.
One can then show that the Ward identity
\begin{eqnarray}\label{eq:NSR_WI}
q_{\mu}\Gamma^{\mu}(P+Q,P)=G^{-1}_0(P+Q)-G^{-1}_0(P)+\Sigma(P)-\Sigma(P+Q)=G^{-1}(P+Q)-G^{-1}(P)
\end{eqnarray}
is satisfied.
Thus the linear response theory based on NSR theory with the vertex function shown in Fig.\ref{fig:NSRvertex} is gauge invariant.

Importantly, this same vertex also satisfies the $Q$-limit Ward Identity
\begin{eqnarray}\label{eq:QWI_NSR}
\lim_{\mathbf{q}\rightarrow \mathbf{0}}\Gamma^{0}(P+Q,P)|_{\omega=0}=1-\frac{\partial\Sigma(P)}{\partial\mu}.
\end{eqnarray}
By explicitly calculating the vertex function, we have
\begin{eqnarray}
\lim_{\mathbf{q}\rightarrow \mathbf{0}}\Gamma^{0}(P+Q,P)|_{\omega=0}=1+\sum_K t_0(K)G^2_0(K-P)-2\sum_{L,K}t^2_0(K)G_0(K-P)G_0(K-L)G^2_0(L).
\end{eqnarray}
Now we evaluate the right hand side of Eq.(\ref{eq:QWI_NSR}) for NSR theory.
\begin{eqnarray}
& &1-\frac{\partial\Sigma(P)}{\partial\mu}=1-\left(\sum_{K}\frac{\partial t_0(K)}{\partial\mu}G_0(K-P)+\sum t_0(K)\frac{\partial G_0(K-P)}{\partial\mu} \right)\nonumber\\
&=&1-2\sum_{L,K}t^2_0(K)G_0(K-P)G_0(K-L)G^2_0(L)+\sum_K t_0(K)G^2_0(K-P)\nonumber\\
&=&\lim_{\mathbf{q}\rightarrow \mathbf{0}}\Gamma^{0}(P+Q,P)|_{\omega=0},
\end{eqnarray}
Thus the $Q$-limit WI is satisfied by the linear
response theory of the NSR theory.
As a consequence, the compressibility sum rule is satisfied by this linear
response theory. We emphasize that these two constraints
(the $Q$-limit Ward identity \eqref{eq:QWI_NSR}
and the Ward identity \eqref{eq:NSR_WI}) are independent constraints \cite{OurPed}.
These observations should be contrasted with
the Hartree-Fock as well as the RPA approximations reviewed in Ref.~\cite{Mahanbook} which cannot reach this level of consistency.

Despite these findings, the compressibility is, nevertheless, ill
behaved, as it diverges when the system approaches $T_c$ from above.
This problem  was
discussed in Ref.~\cite{Lucheroni}, where it was shown that the MT and AL diagrams diverge as $T\rightarrow T_c$.
We note that in NSR theory, $T_c$ is determined by
the temperature where $t_0^{-1}(Q=0)=0$ \cite{NSR},
as in the Thouless criterion in conventional
superconductors \cite{ThoulessCriterion}.
In the number equation, the $t$-matrix resides in the denominator of $G(K)$ so its divergence at $T_c$ does not introduce difficulties.
However, if one calculates $dn/d\mu$ or equivalently calculates the MT and AL diagrams, those expressions explicitly contain
$\frac{d\Sigma}{d\mu}$ in the numerator. In NSR theory, we have
$\frac{d\Sigma}{d\mu}=-\sum_q\frac{d\chi/d\mu}{(1+g\chi)^2}G_0(q-k)+\cdots$. Due to the Thouless criterion, $1+g\chi\to\,aq^2$ for small $q$
when $T\to T_c$. Then the $q$ integral behaves like $\int\frac{q^2dq}{q^4}$ which is divergent at $q=0$.

There have been attempts to remove this divergence by including effective boson-boson interactions \cite{StrinatiPRL12},
but there has been no demonstration that gauge invariance and conservation laws are respected \cite{OurComment} when these repairs are made.
One can identify more generally the problematic aspect of NSR theory. In this
simplest pairing fluctuation scheme, where the $t$-matrix contains only bare Green's functions,
the fermionic chemical potential $\mu$ is effectively the only parameter in the theory.
Once dressed Green's functions are introduced, it becomes possible to
avoid this type of divergence.
At the same time, of course, it becomes more difficult to establish consistency
with gauge invariance and
the compressibility sum rule. Given that the compressibility sum rule
is rarely satisfied (the exceptions being the two examples discussed in
this paper), one has a choice for how to approach a calculation
of the compressibility. We emphasize that from
the literature it appears that the
more credible results for the compressibility arise \cite{Singwi} via
the thermodynamic rather than the two body correlation functions.

\section{Conclusion}
Linear response theories have been an important tool for studying transport and dynamic properties of superfluid and related many-particle systems.
The current focus in the literature
on ultracold Fermi superfluids, particularly at unitarity, provided
a primary motivation for
our
work which aimed
to organize this subject matter and clarify the
constraints that calibrate a linear response theory.
We have seen that
the challenge is to construct such theories as to be fully compatible
with $f$-sum and compressibility-sum rules which reflect conservation
principles. Although there have been some successes there are nevertheless
important failures.

In this paper we presented two nearly unique examples of fermionic superfluids
which are demonstrably consistent with $f$-sum and
compressibility-sum rules.
We addressed these theories via
the compressibility $\kappa$. Important
here was that both are
compatible with the compressibility sum rule. It is useful to compare
these two observations in the normal phase. In effect, the BCS-BEC mean
field approach treated the normal phase as a normal Fermi liquid
and the resulting compressibility is plotted in Figure~\ref{fig:C}.
Above $T_c$ one finds very little temperature dependence.
This should be contrasted with the behavior found in the Nozieres
Schmitt-Rink approach to the normal phase,
where there is a dramatic upturn in the
compressibility with decreasing
temperature. Precisely at $T_c$ the NSR theory predicts that $\kappa$ diverges \cite{Lucheroni}.

One can view the first of these two systems as indicating the
behavior of $\kappa$ associated with a purely fermionic system. By contrast
the dramatic upturn in $\kappa$ with decreasing $T$
is expected for a bosonic system en route to condensation.
Experimentally \cite{Zwierleincompress} the situation for unitary
gases is somewhat in between these two limits. This will be an important
topic for future research.

\ \\
\ \\
Hao Guo thanks the support by National Natural Science Foundation of China (Grants No. 11204032) and Natural Science Foundation of Jiangsu Province, China (SBK201241926). C. C. C. acknowledges the support of the U.S. Department of Energy through the LANL/LDRD Program.
Additional support (KL) is via
NSF-MRSEC Grant
0820054.

\appendix

\section{Detailed expressions for response functions}\label{app:a0}

The following are the EM response functions of fermionic superfluids from the CFOP theory:
\begin{eqnarray}
Q_{11}(\omega,\mathbf{q})&=&\sum_{\mathbf{p}}\Big[\big(1+\frac{\xi^+_{\mathbf{p}}\xi^-_{\mathbf{p}}-\Delta^2}{E^+_{\mathbf{p}}E^-_{\mathbf{p}}}\big)\frac{E^+_{\mathbf{p}}+E^-_{\mathbf{p}}}{\omega^2-(E^+_{\mathbf{p}}+E^-_{\mathbf{p}})^2}[1-f(E^+_{\mathbf{p}})-f(E^-_{\mathbf{p}})] \nonumber \\
& &\quad-\big(1-\frac{\xi^+_{\mathbf{p}}\xi^-_{\mathbf{p}}-\Delta^2}{E^+_{\mathbf{p}}E^-_{\mathbf{p}}}\big)\frac{E^+_{\mathbf{p}}-E^-_{\mathbf{p}}}{\omega^2-(E^+_{\mathbf{p}}-E^-_{\mathbf{p}})^2}[f(E^+_{\mathbf{p}})-f(E^-_{\mathbf{p}})]\Big],
\end{eqnarray}
\begin{eqnarray}
Q_{12}(\omega,\mathbf{q})=-Q_{21}(\omega,\mathbf{q})=-i\omega\sum_{\mathbf{p}}\Big[\big(\frac{\xi^+_{\mathbf{p}}}{E^+_{\mathbf{p}}}+\frac{\xi^-_{\mathbf{p}}}{E^-_{\mathbf{p}}}\big)\frac{1-f(E^+_{\mathbf{p}})-f(E^-_{\mathbf{p}})}{\omega^2-(E^+_{\mathbf{p}}+E^-_{\mathbf{p}})^2}-\big(\frac{\xi^+_{\mathbf{p}}}{E^+_{\mathbf{p}}}-\frac{\xi^-_{\mathbf{p}}}{E^-_{\mathbf{p}}}\big)\frac{f(E^+_{\mathbf{p}})-f(E^-_{\mathbf{p}})}{\omega^2-(E^+_{\mathbf{p}}-E^-_{\mathbf{p}})^2}\Big],
\end{eqnarray}
\begin{eqnarray}
Q^0_{13}(\omega,\mathbf{q})=Q^0_{31}(\omega,\mathbf{q})=\Delta\sum_{\mathbf{p}}\frac{\xi^+_{\mathbf{p}}+\xi^-_{\mathbf{p}}}{E^+_{\mathbf{p}}E^-_{\mathbf{p}}}\Big[\frac{(E^+_{\mathbf{p}}+E^-_{\mathbf{p}})[1-f(E^+_{\mathbf{p}})-f(E^-_{\mathbf{p}})]}{\omega^2-(E^+_{\mathbf{p}}+E^-_{\mathbf{p}})^2}+\frac{(E^+_{\mathbf{p}}-E^-_{\mathbf{p}})[f(E^+_{\mathbf{p}})-f(E^-_{\mathbf{p}})]}{\omega^2-(E^+_{\mathbf{p}}-E^-_{\mathbf{p}})^2}\Big],
\end{eqnarray}
\begin{eqnarray}
\mathbf{Q}^i_{13}(\omega,\mathbf{q})=\mathbf{Q}_{31}^i(\omega,\mathbf{q})=\sum_{\mathbf{p}}\frac{\mathbf{p}^i}{m}\frac{\Delta\omega}{E^+_{\mathbf{p}}E^-_{\mathbf{p}}} \Big[\frac{(E^+_{\mathbf{p}}-E^-_{\mathbf{p}})[1-f(E^+_{\mathbf{p}})-f(E^-_{\mathbf{p}})]}{\omega^2-(E^+_{\mathbf{p}}+E^-_{\mathbf{p}})^2}+\frac{(E^+_{\mathbf{p}}+E^-_{\mathbf{p}})[f(E^+_{\mathbf{p}})-f(E^-_{\mathbf{p}})]}{\omega^2-(E^+_{\mathbf{p}}-E^-_{\mathbf{p}})^2}\Big],
\end{eqnarray}
\begin{eqnarray}
Q_{22}(\omega,\mathbf{q})&=&\sum_{\mathbf{p}}\Big[\big(1+\frac{\xi^+_{\mathbf{p}}\xi^-_{\mathbf{p}}+\Delta^2}{E^+_{\mathbf{p}}E^-_{\mathbf{p}}}\big)\frac{E^+_{\mathbf{p}}+E^-_{\mathbf{p}}}{\omega^2-(E^+_{\mathbf{p}}+E^-_{\mathbf{p}})^2}[1-f(E^+_{\mathbf{p}})-f(E^-_{\mathbf{p}})] \nonumber \\
& &\quad-\big(1-\frac{\xi^+_{\mathbf{p}}\xi^-_{\mathbf{p}}+\Delta^2}{E^+_{\mathbf{p}}E^-_{\mathbf{p}}}\big)\frac{E^+_{\mathbf{p}}-E^-_{\mathbf{p}}}{\omega^2-(E^+_{\mathbf{p}}-E^-_{\mathbf{p}})^2}[f(E^+_{\mathbf{p}})-f(E^-_{\mathbf{p}})]\Big],
\end{eqnarray}
\begin{eqnarray}
Q^0_{23}(\omega,\mathbf{q})=-Q^0_{32}(\omega,\mathbf{q})=i\sum_{\mathbf{p}}\frac{\Delta\omega}{E^+_{\mathbf{p}}E^-_{\mathbf{p}}}\Big[\frac{(E^+_{\mathbf{p}}+E^-_{\mathbf{p}})[1-f(E^+_{\mathbf{p}})-f(E^-_{\mathbf{p}})]}{\omega^2-(E^+_{\mathbf{p}}+E^-_{\mathbf{p}})^2}+\frac{(E^+_{\mathbf{p}}-E^-_{\mathbf{p}})[f(E^+_{\mathbf{p}})-f(E^-_{\mathbf{p}})]}{\omega^2-(E^+_{\mathbf{p}}-E^-_{\mathbf{p}})^2}\Big],
\end{eqnarray}
\begin{eqnarray}
\mathbf{Q}^i_{23}(\omega,\mathbf{q})=-\mathbf{Q}^i_{32}(\omega,\mathbf{q})=i\Delta\sum_{\mathbf{p}}\frac{\mathbf{p}^i}{m}\frac{\xi^+_{\mathbf{p}}-\xi^-_{\mathbf{p}}}{E^+_{\mathbf{p}}E^-_{\mathbf{p}}} \Big[\frac{(E^+_{\mathbf{p}}+E^-_{\mathbf{p}})[1-f(E^+_{\mathbf{p}})-f(E^-_{\mathbf{p}})]}{\omega^2-(E^+_{\mathbf{p}}+E^-_{\mathbf{p}})^2}+\frac{(E^+_{\mathbf{p}}-E^-_{\mathbf{p}})[f(E^+_{\mathbf{p}})-f(E^-_{\mathbf{p}})]}{\omega^2-(E^+_{\mathbf{p}}-E^-_{\mathbf{p}})^2}\Big],
\end{eqnarray}
\begin{eqnarray}
Q_{33}^{00}(\omega,\mathbf{q})&=&\sum_{\mathbf{p}}\Big[\big(1-\frac{\xi^+_{\mathbf{p}}\xi^-_{\mathbf{p}}-\Delta^2}{E^+_{\mathbf{p}}E^-_{\mathbf{p}}}\big)\frac{E^+_{\mathbf{p}}+E^-_{\mathbf{p}}}{\omega^2-(E^+_{\mathbf{p}}+E^-_{\mathbf{p}})^2}[1-f(E^+_{\mathbf{p}})-f(E^-_{\mathbf{p}})] \nonumber \\
& &\quad-\big(1+\frac{\xi^+_{\mathbf{p}}\xi^-_{\mathbf{p}}-\Delta^2}{E^+_{\mathbf{p}}E^-_{\mathbf{p}}}\big)\frac{E^+_{\mathbf{p}}-E^-_{\mathbf{p}}}{\omega^2-(E^+_{\mathbf{p}}-E^-_{\mathbf{p}})^2}[f(E^+_{\mathbf{p}})-f(E^-_{\mathbf{p}})]\Big].
\end{eqnarray}
\begin{eqnarray}
\tensor{Q}_{33}^{ij}(\omega,\mathbf{q})&=&\sum_{\mathbf{p}}\frac{\mathbf{p}^i\mathbf{p}^j}{m^2}\Big[\big(1-\frac{\xi^+_{\mathbf{p}}\xi^-_{\mathbf{p}}+\Delta^2}{E^+_{\mathbf{p}}E^-_{\mathbf{p}}}\big)\frac{E^+_{\mathbf{p}}+E^-_{\mathbf{p}}}{\omega^2-(E^+_{\mathbf{p}}+E^-_{\mathbf{p}})^2}[1-f(E^+_{\mathbf{p}})-f(E^-_{\mathbf{p}})] \nonumber \\
& &\qquad\quad-\big(1+\frac{\xi^+_{\mathbf{p}}\xi^-_{\mathbf{p}}+\Delta^2}{E^+_{\mathbf{p}}E^-_{\mathbf{p}}}\big)\frac{E^+_{\mathbf{p}}-E^-_{\mathbf{p}}}{\omega^2-(E^+_{\mathbf{p}}-E^-_{\mathbf{p}})^2}[f(E^+_{\mathbf{p}})-f(E^-_{\mathbf{p}})]\Big],
\end{eqnarray}
\begin{eqnarray}\label{Q3i}
\mathbf{Q}^{0i}_{33}(\omega,\mathbf{q})=\mathbf{Q}^{i0}_{33}(\omega,\mathbf{q})=\omega\sum_{\mathbf{p}}\frac{\mathbf{p}^i}{m}\Big[\big(\frac{\xi^+_{\mathbf{p}}}{E^+_{\mathbf{p}}}-\frac{\xi^-_{\mathbf{p}}}{E^-_{\mathbf{p}}}\big)\frac{1-f(E^+_{\mathbf{p}})-f(E^-_{\mathbf{p}})}{\omega^2-(E^+_{\mathbf{p}}+E^-_{\mathbf{p}})^2}-\big(\frac{\xi^+_{\mathbf{p}}}{E^+_{\mathbf{p}}}+\frac{\xi^-_{\mathbf{p}}}{E^-_{\mathbf{p}}}\big)\frac{f(E^+_{\mathbf{p}})-f(E^-_{\mathbf{p}})}{\omega^2-(E^+_{\mathbf{p}}-E^-_{\mathbf{p}})^2}\Big].
\end{eqnarray}

\section{Density Structure Factor in the BCS and BEC limits}\label{app:b}
We first consider the BCS limit and
the regime where the external frequency and momentum are small,
such that $0<\omega<2\Delta$
and $0<q\ll k_F$.
Due to the
particle-hole symmetry of strict BCS theory,
$Q_{12}=Q^0_{13}=0$ so $K^{00}=K^{00}_0+\delta K^{00}$, where
$K^{00}_0=Q^{00}_{33}$ and $\delta K^{00}=-Q^0_{23}Q^0_{32}/\tilde{Q}_{22}$.
The
density structure factor is $\chi_{\rho\rho}=\chi_{\rho\rho0}+\delta \chi_{\rho\rho}$,
where $\chi_{\rho\rho0}=-\frac{1}{\pi}\textrm{Im}K^{00}_0$ and
$\delta\chi_{\rho\rho0}=-\frac{1}{\pi}\textrm{Im}\delta K^{00}$ according to Eq.(\ref{dK}).

Our small frequency and small momentum limit guarantees
that it is not possible to break a Cooper pair into two
quasi-particles. Therefore $\chi_{\rho\rho0}$ has no pole. Instead, $\tilde{Q}_{22}$
determines the poles of
$\chi_{\rho\rho}$. This leads to
\begin{equation}\label{eqn:Q22f}
\tilde{Q}_{22}(\omega,\mathbf{q})=-\frac{N(0)}{2\Delta^2}\big(\omega^2-c^2_sq^2\big).
\end{equation}
Here $N(0)$ is the density of states at the Fermi energy. The condition $\tilde{Q}_{22}=0$ yields the excitation dispersion
of the gapless mode
$\omega=c_sq$, where $c_s=\frac{1}{\sqrt{3}}\frac{k_F}{m}$. Similarly, we have
\begin{equation} Q^{0}_{23}(\omega,\mathbf{q})\simeq-\frac{i\Delta\omega
N(0)}{2}\int^{+\infty}_{-\infty}d\xi_{\mathbf{p}}\frac{1}{E^3_{\mathbf{p}}}=-\frac{i\omega
N(0)}{\Delta}. \end{equation}
One then finds for the density structure factor in the BCS limit
\begin{eqnarray} \label{eqn:S0C} \chi_{\rho\rho}(\omega, \mathbf{q})=-\frac{\omega^2
N(0)}{c_sq\pi}\textrm{Im}\Big(\frac{1}{\omega-c_sq+i\delta}-\frac{1}{\omega+c_sq+i\delta}\Big)=\frac{nq}{2mc_s}\delta(\omega-c_sq),
\end{eqnarray} which also satisfies
the $f$-sum rule \begin{eqnarray}\label{eqn:fS2}
\int^{\infty}_0d\omega\omega \chi_{\rho\rho}(\omega,\mathbf{q})=n\frac{q^2}{2m}.
\end{eqnarray}

Next we evaluate the density-density correlation functions in the BEC
limit for different
$\omega, q$ regimes.
We consider three situations
associated with (\textbf{A}) $\frac{\Delta}{|\mu|}\rightarrow 0$ and low frequency
and momentum, (\textbf{B}) $\frac{\Delta}{|\mu|}<1$ and low frequency and momentum,
and (\textbf{C}) $q^2/2m+|\mu|\gg\Delta$ respectively. Case (\textbf{A}) describes
the deep BEC limit where $a\rightarrow0^+$ and $\Delta/|\mu|\rightarrow 0$. Here low
momentum implies $q\ll k_F$ as before while low frequency means $\omega\ll
\Delta_{\textrm{S}}=2\sqrt{|\mu|^2+\Delta^2}$, where $\Delta_{\textrm{S}}$ is the
threshold for fermionic excitations in the BEC regime. Case
(\textbf{B}) corresponds to a relatively shallow BEC regime
as compared to Case
(\textbf{A}). In Case (\textbf{C}), when $q$ is sufficiently large,
the system is in the very
shallow BEC regime where $\mu\rightarrow 0$ at $1/k_Fa=0.553$. This situation was
discussed in Ref.\cite{Stringari06}. The evaluation of the structure factor is
lengthy, but straightforward.

(\textbf{A}). In this case, the system is in the deep BEC limit and can be thought as
a dilute gas of tightly bound
molecules with mass $m_B=2m$. Hence, the gap $\Delta$ is
negligible and we may approximate
$E^{\pm}_{\mathbf{p}}\simeq\xi^{\pm}_{\mathbf{p}}$ and
$\tilde{Q}_{11}\simeq\tilde{Q}_{22}$. Here one finds that
\begin{eqnarray}\label{eqn:Spp0h}
\chi_{\rho\rho}(\omega,\mathbf{q})=\frac{2\Delta^2(2m)^{\frac{3}{2}}\sqrt{|\mu|}}{\pi}\frac{2m}{q^2}\mbox{Arcsin}^2\frac{\frac{q}{\sqrt{2m}}}{\sqrt{\frac{q^2}{2m}+16|\mu|}}\delta(\omega-\frac{q^2}{4m}).
\end{eqnarray}
Note that $\omega=\frac{q^2}{4m}=\frac{q^2}{2m_B}$ appears as an
argument in the delta function, corresponding to
the energy dispersion of free bosons with mass $m_B=2m$. We
note that the fermionic continuum (associated with broken pairs)
does not appear at these low $\omega$.

(\textbf{B}). In this case, the system behaves as
a weakly interacting Bose gas where the
internal structure of the fermion pairs can not be ignored. We expand all response
functions to leading order in $\Delta$, $q^2/2m$ and $\omega$ and assume
$E^{\pm}_{\mathbf{p}}\simeq\xi^{\pm}_{\mathbf{p}}$ with
$\tilde{Q}_{11}\neq\tilde{Q}_{22}$. The density structure factor is given by

\begin{eqnarray}
\chi_{\rho\rho}(\omega,\mathbf{q})=2n\frac{\frac{q^2}{4m}-\frac{\omega^2_s}{8|\mu|}}{\omega_s}\delta(\omega-\omega_s),
\end{eqnarray}
where $\omega_s=\sqrt{c^2_sq^2+(\frac{q^2}{4m})^2}$ is the dispersion of the Bogoliugov
mode.

(\textbf{C}). Since $\Delta\ll q^2/2m+|\mu|$, we expand the energy dispersion relation to
leading order in $\Delta$ as
$E^{\pm}_{\mathbf{p}}=\xi^{\pm}_{\mathbf{p}}+\Delta^2/2\xi^{\pm}_{\mathbf{p}}$\citep{Stringari06}.
If, in addition, the system is in the regime $\frac{\Delta}{|\mu|}<1$, we may
expand the density structure factor to first order in $\Delta^2/|\mu|^2$ to obtain
\begin{eqnarray}
\chi_{\rho\rho}(\omega,\mathbf{q})=2n\big[1-\frac{5}{16}\frac{\Delta^2}{|\mu|^2}-(\frac{1}{24}-\frac{17}{256}\frac{\Delta^2}{|\mu|^2})\frac{q^2}{2m|\mu|}+\cdots\big]\delta(\omega-\omega_c).
\end{eqnarray}

\bibliographystyle{apsrev}

\end{document}